\documentclass[a4paper,aip,jmp,groupedaddress,onecolumn]{revtex4-1}      
\usepackage[utf8]{inputenc}  
\usepackage[T1]{fontenc}     
\usepackage[british]{babel}  
\usepackage[sc,osf]{mathpazo}
\usepackage[scaled=0.86]{berasans}  
\usepackage[colorlinks=true, citecolor=blue, urlcolor=blue]{hyperref}  
\usepackage{graphicx} 
\usepackage[babel]{microtype}  
\usepackage{amsmath,amssymb,amsthm,bm,amsfonts,mathrsfs,bbm} 
\usepackage{xspace}  
\usepackage{pgfplots}  %

\begin{document}
\title{Measurement incompatibility and Schr\"{o}dinger-EPR steering in a class of probabilistic theories}

\author{Manik Banik}
\email{manik11ju@gmail.com}
\affiliation{Physics and Applied Mathematics Unit, Indian Statistical Institute, 203 B. T. Road, Kolkata 700108, India.}


\begin{abstract}
Steering is one of the most counter intuitive non classical features of bipartite quantum system, first noticed by Schr\"{o}dinger at the early days of quantum theory. On the other hand measurement incompatibility is another non classical feature of quantum theory, initially pointed out by N. Bohr. Recently the authors of Refs. [\href{http://journals.aps.org/prl/abstract/10.1103/PhysRevLett.113.160402}{Phys. Rev. Lett. {\bf 113}, 160402 (2014)}] and [\href{http://journals.aps.org/prl/abstract/10.1103/PhysRevLett.113.160403}{Phys. Rev. Lett. {\bf 113}, 160403 (2014)}] have investigated the relation between these two distinct non classical features. They have shown that a set of measurements is not jointly measurable (i.e. incompatible) if and only if they can be used for demonstrating Schr\"{o}dinger-Einstein-Podolsky-Rosen steering. The concept of steering has been generalized for more general abstract tensor product theories rather than just Hilbert space quantum mechanics. In this article we discuss that the notion of measurement incompatibility can be extended for general probability theories. Further we show that the connection between steering and measurement incompatibility holds in a border class of tensor product theories rather than just quantum theory.
\end{abstract}


\maketitle
\section{Introduction}
Quantum theory fundamentally differs in various ways from classical physics. Two of the well studied non classical features of quantum theory are `nonlocality' \cite{Brunner'2014} and `entanglement' \cite{Horodecki'2009}, which involve study of certain peculiar types of correlations among spatially separated multipartite quantum systems. Whereas Bell's theorem \cite{Bell'1964} establishes that there exist quantum correlations that cannot be explained in the framework of \emph{local realistic} world view of classical physics (nonlocality), entanglement reflects the impossibility of describing multipartite quantum systems in convex combination of product states of the involved subsystems \cite{Werner'1989}. Though entanglement is necessary for exhibiting quantum nonlocality, but it is known that these are two distinct concepts \cite{Werner'1989,Brunner'2014,Horodecki'2009}. 

Apart from nonlocality and entanglement, another interesting non classical feature of quantum correlation is `steerability', initially pointed out by Schr\"{o}dinger \cite{Schro'1935,Schro'1936}. To his utter surprise, Schr\"{o}dinger noticed that sharing a bipartite entangled quantum system one can remotely steer the state of the other particle. Though the concept steering is as old as quantum theory, but unlike nonlocality and entanglement it draws attention in very recent time. In ref.\cite{Wiseman'2007}, Wiseman \emph{et al.} have formally introduced the framework of \emph{local hidden state} (LHS) model to study the steering phenomena. They have shown that steering is a different concept from both nonlocality and entanglement, which has been proved in general quantum measurement scenario very recently \cite{Quintino'2015}. Wiseman \emph{et al.} work initiated renewed interest among researcher concerning steering and various important results have been established in last few years \cite{Jones'2007,Branciard'2012,Banik'2014,Moroder'2014,Sun'2014,Quintino'2014,Uola'2014}.

Beside necessity of entanglement, another feature that plays central role in the study of quantum nonlocality is quantum incompatible measurements. Mathematically the notion of incompatibility in quantum theory can be captured in different ways, \emph{eg.} non-commutativity \cite{Varadarajan'1985}, impossibility of joint measurability \cite{Lahti'2003}. Though in the case of projective measurements these two notions are identical, this is not the case for general measurements, i.e., positive-operator-valued-measurements (POVMs). However, in the context of quantum nonlocality it is very natural to capture incompatibility in terms of non joint measurability. It is known that compatible measurements on an (arbitrary) quantum state can never lead to any form of quantum nonlocality. Conversely, in the case of projective measurements, it is known that incompatible measurements can always lead to Bell nonlocality. Considering general binary measurement (two outcome POVMs) it has been shown that measurement incompatibility limits the Clauser-Horne-Shimony-Holt (CHSH) inequality \cite{CHSH'1969} violation in quantum theory \cite{Banik'2013}. On the other hand Wolf \emph{et al.} have shown that any set of two incompatible POVMs with binary outcomes can always lead to violation of the CHSH inequality \cite{Wolf'2009}. Note that pairwise joint measurability in general does not imply full joint measurability of arbitrary number of POVMs with arbitrarily many outcomes \cite{Kraus'1983}. Recently, in \cite{Uola'2014}, it has been shown that the result of Wolf \emph{et al.} cannot be extended to the general case. 

However in this general scenario the authors of \cite{Quintino'2014} and \cite{Uola'2014} have independently established an connection between the incompatibility of quantum measurements and the weaker form of quantum nonlocality, i.e., quantum steering. They have shown that for any set of POVMs (arbitrary in numbers with arbitrary many outcomes) that are incompatible (i.e., not jointly measurable), one can find an entangled state, such that the resulting statistics violates a steering inequality. Hence, the use of incompatible measurements is a necessary and sufficient ingredient for demonstrating EPR steering. 

The concepts of steering has been extended for more general class of theories \cite{Barnum'2010,Barnum'2012,Stevens'2013} known as convex operational theory or generalized probability theory (GPT) \cite{Barrett'2007}. In this work we show that one can introduce the notion of measurement incompatibility in these border class of theories. We further show that the connection between measurement incompatibility and steering as established in \cite{Quintino'2014,Uola'2014} also holds in these broader class of theories rather than just quantum theory. 

The organization of the paper is as following: in section-\ref{sec2} we describe the framework for convex operational theories where we have discuss about the structure of state space, the properties of observable, concept of joint measurement, state space of composite system, concepts of entanglement and steering in these generalized framework, and the concept of marginal state. In section-\ref{sec3} we prove our result, i.e., the connection between measurement incompatibility and steering in the general framework and then we present our conclusion in section-\ref{sec4}. 
\section{Convex operational theories}\label{sec2}
The advent of quantum information theory has been accompanied by an upturn of interest in the convex framework for operational theory. Researchers seek to understand the nature of information processing in increasingly abstract terms as it illuminates the difference between the information processing power of quantum theory and that of classical theory. Furthermore, it renews interests in foundational aspects of quantum theory, often with new twists in the axioms or principles concerning information processing are considered. The framework was initially introduced in the 1960's by researchers in quantum foundations who used it to investigate axiomatic derivations of the Hilbert space formalism of quantum mechanics from operational postulates \cite{Mackey'1963,Ludwig'1967,Ludwig'1968,Davies'1970}. Due to the emphasis on the convex structure of the set of states and the use of operations to model state transformations, the approach is called convex state approach. The basic motive of this framework is to explain the experimental phenomena in an operational approach. So the theories considered in this framework are also specified under a common name, called operational theories. Recently, the framework has gained renewed interest from researchers in quantum information science exploring the information theoretic foundations of quantum mechanics. The theories encapsulated in this framework are also known as generalized probabilistic theories (GPT's) \cite{Barrett'2007,Janotta'2011,Barnum'2009,Hardy'2011,Janotta'2014}. In the following we explicitly describe the mathematical framework.
\subsection{State space}
In generalized probabilistic theories, the set of states $\Omega$, in which a system S can be prepared in, is commonly assumed to be a convex subset of a real vector space $V$. The convexity corresponds to the ability to define a preparation procedure as a probabilistic mixture of preparation procedures corresponding to other states, i.e., for every two elements $\omega_1,\omega_2\in\Omega$, their convex combination $C_{\omega_1,\omega_2}:=\{p\omega_1+(1-p)\omega_2|0\le p\le 1\}$ is contained in $\Omega$. The convexity of the state space $\Omega$ can be expressed in more general way. For any set of states $\{\omega_k\}_k\subset\Omega$ with respect to probabilities $\{p_k\}_k$ the convex mixture is defined as:
\begin{equation}\label{eq1.1}
\omega=\sum_kp_k\omega_k,~~~\mbox{where}~~\omega_k\in\Omega~~\forall~~k~~\mbox{and}~~\sum_kp_k=1.
\end{equation}
The convexity of $\Omega$ demands $\omega\in\Omega$. However, the requirement that convex sums of arbitrarily (but finitely) many elements of the set $\Omega$ have to be contained in $\Omega$ can be reduced to the requirement that the convex sum of only two elements has to be contained in $\Omega$. The extremal points of $\Omega$ are refereed as \emph{pure} states and the states which can be written as convex combinations of other states are called \emph{mixed} states. If the number of extremal points are finite then the convex set is called polytopes and a special type of polytopes is simplex where the mixed states have unique decomposition in terms of extremal points. Classical probability theory is simplectic. On the other hand quantum theory is neither simplectic nor polytopic but a convex set where the state space is given by the set of density operators, denoted as $\mathcal{D}(\mathcal{H}_S)$, acting on the Hilbert space $\mathcal{H}_S$ associated with a quantum system $S$. Extremal states $\rho\in\mathcal{D}(\mathcal{H}_S)$ are characterized by $\mbox{Tr}(\rho^2)=\mbox{Tr}(\rho)=1$.  
\subsection{Observable}
The abstract state space, introduced above, in turn gives rise to the mathematical structure of measurements. The set of affine functionals on $\Omega$ forms an ordered linear space $\mathcal{A}(\Omega)$, with the ordering given point-wise: $\mathcal{A}(\Omega)\ni f\ge 0$ if $f(\omega)\ge 0$ for all $\omega\in\Omega$. $\mathcal{A}(\Omega)$ is an order unit space, with order unit defined as : $u$ such that $u(\omega)=1$ for all $\omega\in\Omega$. The set of effects on $\Omega$ is taken to be the unit interval $[0, u]\subset \mathcal{A}(\Omega)$ which is denoted as:
\begin{equation*}
\mathcal{E}(\Omega):=\{e\in A(\Omega)~|~0\le e(\omega)\le 1,~\forall~\omega\in\Omega\}.
\end{equation*}
$\mathcal{E}(\Omega)$ is the convex hull of the unit effect, the zero effect and a set of extremal effects and is a subset of the vector space $V^*$, which is dual to the vector space $V$. In this convex framework, one can, however, define unnormalized states as well as unnormalized effects. The collection of unnormalized states forms a convex positive cone lying in $V^{+}$. Similarly, collection of unnormalized effects from the corresponding dual positive cone lying in $V^{_*+}$.
A discrete observable $\mathcal{O}$ is then a function from an outcome set $\mathcal{K}$ into $\mathcal{E}(\Omega)$ satisfying the normalization condition, i.e., every outcome $k\in\mathcal{K}$ corresponds to an effect $e^{k}\in\mathcal{E}(\Omega)$ such that $\sum_{k\in\mathcal{K}}e^{k}=u$. In quantum theory observables are positive-operator-valued-measurement (POVM): $\{E^k~|~E^k\ge 0,\sum_kE^k=\mathbf{1}\}$ where the POVM element $E^k$ corresponds to the effect corresponding to outcome $k$.  

Besides identifying the space of states and operators, a theory must assigns a rule to calculate the outcome probability of any measurement, $p(e^{k}|\omega)\equiv e^k(\omega):\Omega\times\mathcal{A}(\Omega)\rightarrow[0,1]$. The value $p(e^{k}|\omega)$ denotes the probability of getting outcome $k$ for a measurement of the observable $\mathcal{O}$ in state $\omega$. In quantum theory this outcome probability is given by the generalized Born rule, $\mbox{Tr}(\rho E_k)$. However for our purpose in this work we do not require the explicit mathematical form of states and effects and the explicit rule giving the outcome probability.

Convex combinations of the effects are again a valid effect, i.e., for any $\{e^1,e^2,...,e^r\}\subset\mathcal{E}(\Omega)$ and a probability distribution $\{p_i\}_{i=1}^r$, $e=\sum_ip_ie^i\in\mathcal{E}(\Omega)$; and the outcome probability of the effect $e$ on any state $\omega$ is given by $e(\omega):=p_1e^{1}(\omega)+p_2e^{2}(\omega)+...+p_re^{r}(\omega)$. It is the convexity of states and effects for which this framework is called convex operational framework. 
\subsection{Joint measurement}
A set of observables is said to be jointly measurable if all of them can be evaluated in a single measurement, meaning that there exists a measurement apparatus that contains all the effects of these observables as marginals. A GPT may also include effects that cannot be measured jointly. Therefore, it is of interest to formulate a general criterion for joint measurability. A set of $m$ observables $\mathcal{O}_j\equiv\{e^{k_j}\}$ is called jointly measurable if there exists a measurement $\{e^{\vec{k}}\}$ with outcome $\vec{k}=[k_{j=1},k_{j=2}, . . . ,k_{j=m}]$ where $k_j$ gives the outcome of $j^{th}$ measurement, i.e.,
\begin{equation}
e^{\vec{k}}\ge 0,~~\sum_{\vec{k}}e^{\vec{k}}=u,~~\sum_{\vec{k}\backslash k_j}e^{\vec{k}}=e^{k_j}~\forall~j.\label{joint-meas}
\end{equation}
where $\vec{k}\backslash k_j$ stands for the elements of $\vec{k}$ except for $k_j$. Hence, all effects $e^{k_j}$ are recovered as marginals of the mother observable $e^{\vec{k}}$. Important to note that if $m$ observables are jointly measurable then any subset of these $m$ observables are also jointly measurable. However the converse is not true in general, i.e., joint measurability of all possible proper subsets of these $m$ observables does not necessarily imply that they are jointly measurable in all together. 
\subsection{State space for composite system}
Suppose systems $A$ and $B$ have state spaces $\Omega_A$ and $\Omega_B$. The joint system $AB$ will have its own state space, $\Omega_{AB}$, which is convex by definition. Naturally the question arise: how are $\Omega_{A}$, $\Omega_{B}$, and $\Omega_{AB}$ related? In general, one can imagine many weird and wonderful relations among these state spaces \cite{Namioka'1969}. However, one can narrow down these possibilities significantly by imposing the following quite natural conditions:
\begin{itemize}
\item[(1)] a joint state defines a joint probability for each pair of effects $(e_A, e_B)$, where $e_A\in \mathcal{E}(\Omega_A)$ and $e_B\in \mathcal{E}(\Omega_B)$;
\item[(2)] these joint probabilities respect the no-signaling principle; i.e., the marginal probabilities for the outcomes of a measurement on $B$ do not depend on which measurement was performed on $A$ and vice versa;
\item[(3)] if the joint probabilities for all pairs of effects $(e_A, e_B)$ are specified, then the joint state is specified. This condition is known as \emph{local tomography} condition \cite{Hardy'2011}.
\end{itemize}
These conditions ensure that the linear space $V_{AB}$ in which the joint state space $\Omega_{AB}$ and the cone of associated unnormalized states are embedded can be taken to be $V_A\otimes V_B$. Furthermore, it must lie between two
extremes, the \emph{maximal} and the \emph{minimal} tensor products, defined as:

\begin{itemize}
\item \emph{Maximal tensor product}: Denoted as $\Omega_A\otimes_{max}\Omega_B$, is the set of all
bilinear functionals $\phi:~V^*_A\otimes V^*_B\rightarrow \mathbb{R}$ such that (i) $\phi(e_A,e_B)\ge 0$ for all $e_A\in \mathcal{E}(\Omega_A)$ and $e_B\in \mathcal{E}(\Omega_B)$ and (ii) $\phi(u_A,u_B)=1$, where $u_A$ and $u_B$ are unit effects for system $A$ and $B$ respectively. The maximal tensor product has an important operational characterization: it is the largest set of states assigning probabilities to all product measurements but not allowing signaling.

\item \emph{Minimal tensor product}: Denoted as $\Omega_A\otimes_{min}\Omega_B$, is the convex hull of the product states, where a product $\omega_A\otimes\omega_B$ is defined by $(\omega_A\otimes\omega_B)(a,b)=\omega_A(a)\omega_B(b)$ for all pairs of $(a,b)\in V^*_A\otimes V^*_B$.
\end{itemize}

\emph{Product, separable and entangled states:} If these systems are completely independent, their joint state is given by a product state, denoted as $\omega_{AB}=\omega_A\otimes\omega_B$. Similarly, the effects of the two subsystems can be combined in product effects $e_{AB}=e_{A}\otimes e_B$, describing statistically independent measurements on both sides. In this situation the joint measurement probabilities factorize, i.e.,
$e_{AB}(\omega_{AB})= p(e_{A}\otimes e_B|\omega_A\otimes\omega_B)
= p(e_A|\omega_A)p(e_B|\omega_B)
= e_A(\omega_A)e_B(\omega_B)$.
However not all states and not all effects are of the product form. A bipartite state will be called separable or classically correlated if it can be expressed as probabilistic mixture of product states, i.e.,
\begin{equation}
\omega^{sep}_{AB}=\sum_ip_i\omega^i_A\otimes\omega_B^i.\label{sep}
\end{equation} 
In the GPT framework the set which yields the set of product elements and their probabilistic mixtures, is the so-called the minimal tensor product. For classical systems one can show that the minimal tensor product already includes all joint elements that are consistent with the division into classical subsystems. However, in non-classical theories there are generally additional states which are non-separable but nevertheless consistent with the subsystem structure and fully identifiable by local operations and classical communication (LOCC). Such states are called entangled state.

\subsection{Marginal and conditional states}
Let us consider \emph{local} measurement performed on arbitrary bipartite state $\omega_{AB}\in\Omega_{AB}$. Since the local measurements are independent, we do not have to apply the effects $e^k_{A}$ and $e^l_{B}$ at once. In particular, we could observe only the outcome in part A, ignoring the measurement in part B. The probability of this outcome is given by the marginal probability 
\begin{eqnarray}
p(e^k_A|\omega_{AB})&=&\sum_lp(e^k_A, e^l_B|\omega_{AB})\nonumber\\
&=&\sum_l[e^k_A\otimes e^l_B](\omega_{AB})\nonumber\\
&=& e^k_A\otimes \left[\sum_le^l_B\right] (\omega_{AB})\nonumber\\
&=&[e^k_A\otimes u_B](\omega_{AB})\nonumber\\
&=& e^k_A(\omega_A^{u_B}).
\end{eqnarray}
Here $\omega_A^{u_B}\in\Omega_A$ is called the marginal state of the system $A$, given that the composite system's state is $\omega_{AB}$. The marginal state $\omega_A^{u_B}$ reflects our knowledge about subsystem $A$ provided that potential measurements on subsystem $B$ are ignored. However, our knowledge is of course different if a particular measurement on $B$ is carried out and the result is known to us via classical communication. This increased knowledge is accounted for by the conditional probabilities
\begin{eqnarray}
p(e^k_A|e^l_B,\omega_{AB})&=&\frac{p(e^k_A, e^l_B|\omega_{AB})}{p(e^l_B|\omega_{AB})}\nonumber\\
&=&\frac{[e^k_A\otimes e^l_B](\omega_{AB})}{e^l_B(\omega_B^{u_A})}\nonumber\\
&=& e^k_A\left(\frac{\tilde{\omega}_A^{e^l_B}}{e^l_B(\omega_B^{u_A})}\right)\nonumber\\
&=& e^k_A(\omega_A^{e^l_B}).
\end{eqnarray}
$\omega_A^{e^l_B}$ denotes normalized conditional state on Alice's side and tilde denotes the unnormalized version.

\subsection{Steering in GPT}
The concept of steering was initially pointed out by Schr\"{o}dinger \cite{Schro'1935,Schro'1936}, and he observed that all bipartite pure entangled states are steerable. Likewise the local hidden variable (LHV) model in Bell scenario, Wiseman \emph{et al.} have introduced the framework of \emph{local hidden state} (LHS) model to study the phenomena of steering \cite{Wiseman'2007}. They have shown that steering is not only a specific feature of pure entangled states as noticed by Schr\"{o}dinger, but there exist mixed entangled states which are steerable. It has been also proved that the concept of steering is different from both the concepts nonlocality and entanglement \cite{Wiseman'2007,Quintino'2015}. Whereas nonlocality and entanglement are symmetric concepts, the concept of steering is inherently asymmetric. Note that the example of one way steerable states \cite{Midgley'2010,Handchen'2012,Bowles'2014} have been reported recently. Note that the concept of steering can easily be extended for more general probability theory.

Let Alice and Bob share a bipartite state $\omega_{AB}\in\Omega_{AB}$. Alice choses her measurement from the set $\{e^{k_x}\}_x$, where the index $x$ denotes measurement choice and the index $k_x$ denotes measurement outcome. Upon performing measurement
$x$, and obtaining outcome $k_x$, the sub-normalized state held by Bob is given by 
\begin{equation}
\tilde{\omega}_B^{e^{k_x}}=[e^{k_x}\otimes u_B](\omega_{AB}).
\end{equation}
The sub-normalization condition implies that $0\le u_B(\tilde{\omega}_B^{e^{k_x}})\le 1$. The set of sub-normalized states $\{\tilde{\omega}_B^{e^{k_x}}\}$ is referred as an assemblage for the unconditional marginal state $\omega_B^{u_A}\in\Omega_B$ of Bob's system. No signaling (from Alice to Bob) condition is satisfied as $\sum_x\tilde{\omega}_B^{e^{k_x}}=\omega_B^{u_A}=\sum_{x'}\tilde{\omega}_B^{e^{k_{x'}}}$, for all choice of measurements $x$ and $x'$.

In a steering test, Alice want to convince Bob that the state $\omega_{AB}$ is entangled, i.e., not of the form as given in Eq.(\ref{sep}). Bob does not trust Alice, and thus wants to verify Alice's claim. Asking Alice to perform a given measurement $x$, and to announce the outcome $k_x$, Bob can determine the assemblage $\{\tilde{\omega}_B^{e^{k_x}}\}$ by performing the experiment large number of times. To ensure that steering did indeed occur, Bob should verify that the assemblage does not admit a decomposition (LHS model) of the form
\begin{equation}
\tilde{\omega}_B^{e^{k_x}}=\sum_{\lambda}\Gamma(\lambda)p(k|x,\lambda)\omega^{\lambda},~~\forall~k_x,\label{unsteer}
\end{equation}
where $\sum_{\lambda}\Gamma(\lambda)= 1$. Clearly, if a decomposition of the above form exists, then Alice could have cheated by sending the \emph{local} state $\omega^{\lambda}$ to Bob and announce outcome $k_x$ to Bob according to the distribution $p(k|x,\lambda)$. Note that here $\lambda$ represents a local variable of Alice, representing her choice of strategy.

Assemblages of the form (\ref{unsteer}) are called `un-steerable' and an assemblage which does not satisfy such decomposition is called `steerable'. 
\begin{itemize}
\item[]\emph{Definition 1:} A state $\omega_{AB}\in\Omega_{AB}$ is called steerable for its $B$ marginal if there exists at least one steerable decomposition of that marginal which Alice can remotely prepare by performing local measurement on her particle.
\end{itemize}
The marginal of $B$ may have many more (even uncountably many) steerable decompositions. A state $\omega_{AB}$ is steerable for its $B$ marginal does not necessarily imply that all such decompositions can be prepared by Alice by performing local measurement on her side. So we introduce the concept of \emph{strongly} steerable state.

\begin{itemize}
\item[]\emph{Definition 2:} A state $\omega_{AB}\in\Omega_{AB}$ is called strongly steerable for its $B$ marginal if all possible decompositions (both steerable and un-steerable) of that marginal can be remotely prepared by Alice by performing local measurement on her particle.
\end{itemize}

As for example, according to Gisin-Hughston-Jozsa-Wootters (GHJW) theorem \cite{Gisin'1989,Hughston'1993}, bipartite pure entangled quantum states are strongly steerable whereas there are mixed entangled states which are steerable but not strongly steerable as in the above sense.

\begin{itemize}
\item[]\emph{Definition 3:} A general probabilistic model of a system $B$ with state space $\Omega_B$ supports uniform universal steering if there is another system $A$ with state space $\Omega_A$, such that for any $\omega_B\in\Omega_B$, there is a state $\omega_{AB}\in\Omega_{AB}$, with $\omega_B^{u_A}=\omega_B$ that is steering for its $B$ marginal, and supports universal self-steering if the above is satisfied with $B = A$.
\end{itemize}

\section{Measurement incompatibility \& steering}\label{sec3}
We are now in a position to prove our main result. Recently, the authors of Ref.\cite{Quintino'2014} and Ref.\cite{Uola'2014} have, independently,  established that non joint measurement of quantum POVMs and steerability of bipartite entangled quantum states are equivalent concept. More precisely they have shown that a set of quantum measurements is not jointly measurable if and only if it can be used for demonstrating Einstein-Podolsky-Rosen steering, a form of quantum nonlocality.  In the following we show that this equivalence holds not only in quantum theory but for any no-signaling theory which allows strong steerability.

\begin{itemize}
\item[]{\bf Theorem:} The assemblages $\{\tilde{\omega}_B^{e^{k_x}}\}$, with $\tilde{\omega}_B^{e^{k_x}}=[e^{k_x}\otimes u_B](\omega_{AB})$, is un-steerable for any state $\omega_{AB}\in\Omega_{AB}$ if and only if the set of effects $\{e^{k_x}\}\subset\mathcal{E}(\Omega_A)$ is jointly measurable.
\end{itemize}

{\bf Proof:}
\emph{`if' part}$\rightarrow$ Here our aim is to show that $\{\tilde{\omega}_B^{e^{k_x}}=[e^{k_x}\otimes u_B](\omega_{AB})\}$, for any state $\omega_{AB}\in\Omega_{AB}$, admits a decomposition of the form (\ref{unsteer}) when the set $\{e^{k_x}\}_x$ is jointly measurable.

As the set $\{e^{k_x}\}_x$ is jointly measurable, then according to condition (\ref{joint-meas}) there exists a mother observable such that all the effects in the set $\{e^{k_x}\}_x$ are reproduced as marginal of that mother observable.

Let $e^{\vec{k}}$ be the mother observable with $\vec{k}=[k_{x=1},k_{x=2},...]$ and 
\begin{equation*}
e^{\vec{k}}\ge 0,~~\sum_{\vec{k}}e^{\vec{k}}=u_A,~~\sum_{\vec{k}\backslash k_x}e^{\vec{k}}=e^{k_x}.
\end{equation*}
Define Alice's local variable to be $\lambda=\vec{k}$, distributed according to $\Gamma(\vec{k})=e^{\vec{k}}(\omega^{u_{_B}}_A)$, where $\omega^{u_{_B}}_A$ is Alice's marginal of the bipartite state $\omega_{AB}$. Next Alice sends the local state $\omega^{\vec{k}} = [e^{\vec{k}}\otimes u_B](\omega_{AB})/\Gamma(\vec{k})$. When asked by Bob to perform measurement $x$, Alice announces an outcome $k$ according to $p(k|x,\vec{k})=\delta_{k,k_x}$.

\emph{`only if' part}$\rightarrow$ Consider an arbitrary state $\omega_{AB}\in\Omega_{AB}$. The assemblage resulting from a set of observables $\{e^{k_x}\}$ on state $\omega_{AB}$ is given by 
\begin{equation*}
\tilde{\omega}_B^{e^{k_x}}=[e^{k_x}\otimes u_B](\omega_{AB}).
\end{equation*} 
Our aim is to show that if $\tilde{\omega}_B^{e^{k_x}}$ is un-steerable then $\{e^{k_x}\}$ is jointly measurable, i.e., there exists a mother observable $e^{\vec{k}}$ which gives $\{e^{k_x}\}$ as marginals. As $\tilde{\omega}_B^{e^{k_x}}$ is un-steerable, we have that
\begin{equation*}
\tilde{\omega}_B^{e^{k_x}}=\sum_{\lambda}\Gamma(\lambda)p(k|x,\lambda)\omega^{\lambda},
\end{equation*}
where $\sum_{\lambda}\Gamma(\lambda)=1$ and $\omega^{\lambda}\in\Omega_A$. Let $e^{\lambda}$ be the effect on Alice's side such that   
\begin{equation*}
\omega^{\lambda}\sim\tilde{\omega}_B^{e^{\lambda}}=[e^{\lambda}\otimes u_B](\omega_{AB}).
\end{equation*}
Let us define the effect $e^{\vec{k}}$ as following
\begin{equation*}
e^{\vec{k}}:=\sum_{\lambda}\Gamma(\lambda)e^{\lambda}\prod_xp(k_x|x,\lambda).
\end{equation*}
Clearly we have
\begin{eqnarray*}
\left[\sum_{\vec{k}\setminus k_x}e^{\vec{k}}\otimes u_B \right](\omega_{AB})&=&  \sum_{\lambda}\Gamma(\lambda)p(k|x,\lambda)\omega^{\lambda}\\
&=& [e^{k_x}\otimes u_B](\omega_{AB}).
\end{eqnarray*}
Therefore $e^{\vec{k}}$ is the mother effect which produces $\{e^{k_x}\}$ as marginals.

Note that, sharing a strongly steerable state $\omega_{AB}$ Alice can remotely prepares all possible decomposition of $B$ marginal by performing measurement on her part of $\omega_{AB}$. Whenever a set of measurements is incompatible, i.e., not jointly measurable then performing those measurements Alice produces steerable decomposition of Bob's marginal state. 

At this point it should be noted that the above connection does hold in any tensor product theories. As for example if we consider minimal tensor product structure then it does not allow strongly steerable state and hence the connection will not hold is such case.

\section{Discussion and concluding remarks}\label{sec4}
The concept of steering is nearly as old as quantum theory. Though this concept is very much disturbing to accept, but it has no contradiction with relativistic causality. It does not directly implies Bell's nonlocality, rather it is an weaker form of nonlocality than Bell's nonlocality. On the other hand measurement incompatibility (i.e. non joint measurability) is another important feature of quantum theory which makes it different from classical physics. Though measurement incompatibility plays important role in Bell's nonlocality, but recently it has been proved that measurement incompatibility is directly connected with weaker form of nonlocality, i.e., steering \cite{Quintino'2014,Uola'2014}. 

The concept of measurement incompatibility and the concept of steering can be extended in more general class of probability theories with quantum theory a special example of this class. Here we show that the connection between measurement incompatibility and steering as established in \cite{Quintino'2014,Uola'2014} also holds in a broader class of theories allowing strongly steerable state.   

{\bf Acknowledgment:} Author gratefully acknowledges private communications with Nicolas Brunner.

\end{document}